\documentstyle[12pt]{article}    
\input epsf  
\newcommand{\mw}{\mbox{$\overline {m \omega}$}}
\renewcommand{\theequation} {{\arabic{section}}.{\arabic{equation}}}
\begin{document}                    

\centerline{\Large{\bf   Algebraic Models:}}
\centerline{\Large{\bf Coordinates, Scales, and Dynamical Symmetries}}

\vspace{.2in}
\centerline{Michael W.N. Ibrahim}

\vspace{.2in}
\centerline{\em Center for Theoretical Physics,
                Sloane Laboratory,}
\centerline{\em Yale University, 
                                New Haven, Connecticut\ \ 06520--8120}

\vspace{.5in} \centerline{\bf Abstract} 

We discuss the variety of coordinates often used to characterize the
coherent state classical limit of an algebraic model.  We show
selection of appropriate coordinates naturally motivates a procedure
to generate a single particle Schr\"odinger hamiltonian which, for low
energy states, gives equivalent results to a bosonic algebraic model
to leading order in $N$.  The process is used to study the associated
geometries of the dynamical symmetries of $U(3)$.  By demanding that
the inner product be preserved in the Schr\"odinger picture we
conclude that different dynamical symmetries correspond to different
scales.
\vspace{.5in}


\section{Introduction} \setcounter{equation} {0}

Recently, hybrid algebraic-Schr\"odinger approaches for the study of
transition intensities in molecules have been introduced
[\ref{II}--\ref{IL}].  In these approaches one relies on
Schr\"odinger/algebraic correspondences to introduce operators in the
algebra which are functions of configuration space parameters. For the
$U(2)$ model considered in [\ref{II},\ref{MDVPII}] one does so by
relying on correspondences between a particular dynamical symmetry and
a Schr\"odinger Hamiltonian [\ref{MANY}].  For more complicated
algebras such correspondences are not available.  Further, the
inner-product of two algebraic states is not equal to the
Schr\"odinger inner-product of their images.  Since, in some
circumstances, the inner-product of the algebra has a natural
interpretation as a Schr\"odinger overlap [\ref{ME}] such shortcomings
are an impediment to understanding molecular structure.

It would therefore be useful to have a generic correspondence
---independent of a particular dynamical symmetry--- for the
interpretation of algebraic parameters as geometric quantities which
reside in configuration space (which is noticeably absent in the
algebraic models).  We provide such an interpretation by exploiting
the many approximate correspondences between algebraic and
Schr\"odinger pictures.  We demonstrate how one may turn an algebraic
hamiltonian into a traditional Schr\"odinger (traditional meaning
kinetic plus potential---with no coordinate dependent mass terms)
single particle hamiltonian which will give the same results to
leading order in $N$, the label for the symmetric representation of
the algebra.  Our procedure is most similar to the re-quantization used
by [\ref{ROOS}] to study spectra. We proceed further to study the
wavefunctions and inner product.  Our approach differs from other
studies which have examined the correspondence between algebraic and
Schr\"odinger inner products [\ref{WULFMAN}] since they rely on
particular dynamical symmetries. We work out our procedure explicitly
for $U(3)$ and conclude that different chains correspond not only to
different geometries but different scales.

\section{Algebraic Models: Constructing a Connection with Configuration Space} \setcounter{equation} {0}

Numerous correspondences between algebraic and Schr\"odinger methods and their classical limits can be succinctly summarized in the following simple diagram:
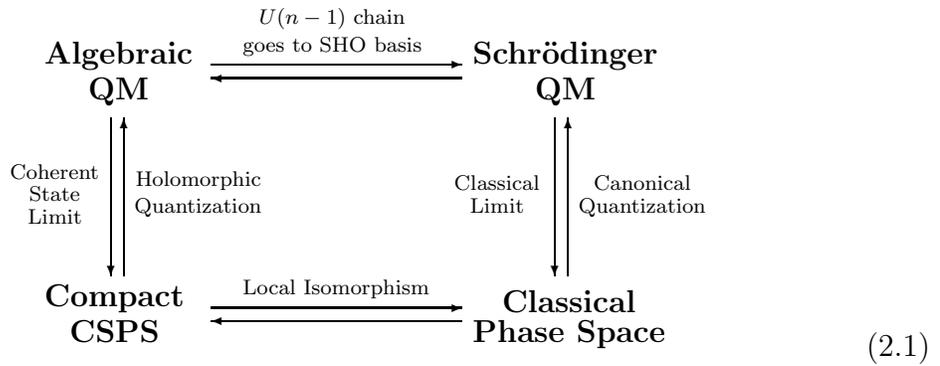
\begin{equation}
\label{eqn:correspondence}
\begin{picture}(250,140)
\put(23,97){\shortstack{\bf Algebraic \\ \bf QM}}
   \put(86,100){\begin{picture}(75,50)
                \put(12,15){\shortstack{\scriptsize $U(n-1)$ chain \\
                                       \scriptsize  goes to SHO basis}}
                \put(0,10){\vector(1,0){95}} 
                \put(95,5){\vector(-1,0){95}}
                \end{picture}}
\put(185,97){\shortstack{ \bf Schr\"odinger \\ \bf QM}}
   \put(10,20){\begin{picture}(50,100)
       \put(0,30){\shortstack{\scriptsize Coherent  \\
                                      \scriptsize State \\
                                      \scriptsize Limit}}
               \put(47,34){\shortstack{\scriptsize Holomorphic \\
                                      \scriptsize Quantization}}
               \put(43,10){\vector(0,1){60}}
               \put(38,70){\vector(0,-1){60}}
               \end{picture}}
\put(23,5){\shortstack{\bf  Compact\\ \bf  CSPS}} 
   \put(178,20){\begin{picture}(50,100)
       \put(0,34){\shortstack{\scriptsize Classical \\
                              \scriptsize Limit}}
               \put(47,34){\shortstack{\scriptsize Canonical \\
                                      \scriptsize Quantization}}
               \put(43,10){\vector(0,1){60}}
               \put(38,70){\vector(0,-1){60}}
               \end{picture}}
   \put(86,8){\begin{picture}(75,50)
                \put(12,15){\scriptsize Local Isomorphism}
                \put(0,10){\vector(1,0){95}} 
                \put(95,5){\vector(-1,0){95}}
                \end{picture}} 
\put(185,5){\shortstack{\bf Classical \\ \bf  Phase Space}}
\end{picture}
\end{equation}
Hamiltonians on the right hand side are written in terms of geometric
parameters whereas those on the left are written in terms of algebraic
ones.  Thus, a careful exploitation of the horizontal correspondences
should relate the different parameters.  We propose to do so by travelling
counter-clockwise around the diagram---re-quantizing the algebraic model.

The vertical correspondences are extremely well defined and have been
thoroughly discussed in the literature
[\ref{ROOS},\ref{OriginalGilmore}--\ref{ZFG}].  The relationship
between the $U(n) \supset U(n-1)$ and the simple harmonic oscillator
(SHO) has also been discussed [\ref{IA}] and can be formalized by
considering the contraction limit of the algebra
[\ref{GILMOREBOOK},\ref{IL}].  Similarly, if one finds appropriate
coordinates in a patch of the coherent state phase space (CSPS) one
may embed this region within a standard phase space and have a perfect
copy of the dynamics for local trajectories [\ref{A}].

The horizontal correspondences are approximate on the whole given that the
phase spaces are topologically different (compact versus non-compact)
in the classical regimes and one has a finite versus infinite
dimensional Hilbert space in the corresponding quantum cases.
However, with careful selection of coordinates it will be seen
that these incompatabilities will not effect the lower bound states in a
re-quantization scheme.

It is to the selection of these coordinates we now turn. We must find
coordinates in a patch of CSPS which behave like positions and momenta
and then identify these as coordinates covering the entire canonical
phase space.  In the language of geometry, we wish to find coordinates
in which the symplectic form determining the dynamics is in Darboux
form: $\omega = dp \wedge dq$ [\ref{A}].  Historically, it has been
useful to think of the imaginary part of the coherent state parameter
$\alpha$ as a momenta [\ref{IA},\ref{ROOS}].  We adopt that convention
here to eliminate any remaining coordinate choice ambiguity.

Examining the action in the propagator path integral
[\ref{ZFG},\ref{ROOS}]: one finds that the classical hamiltonian is
given by the coherent state expectation, $H = \langle \alpha |\hat H|
\alpha \rangle $; $|\alpha \rangle = \frac{1}{\sqrt{N!}}(
\sqrt{1-\alpha \cdot \alpha^*} \sigma^{\dagger} + \alpha \cdot
\tau^{\dagger})^N|0 \rangle $ denotes the coherent state in group
coordinates [\ref{OriginalGilmore}]; $\sigma$ and $\tau$ are typically
scalar and tensor operators of $O(3)$ respectively; and the hamiltonian evolution is determined by the
symplectic form $\omega = i \hbar d \langle \alpha |d|\alpha \rangle $
which in group coordinates is explicitly:
\begin{eqnarray}
\omega = N i \hbar \, d \alpha_{\mu}^* \wedge  d \alpha_{\mu}.
\end{eqnarray}

Deviating slightly from the typical development we let
\begin{eqnarray}
\label{eqn:alphascale}
 \alpha_{\mu} = {\sqrt \frac{\mw}{2}} (\frac{q_{\mu}}{\sqrt N} +
\frac{i}{\mw} \frac{p_{\mu}}{\sqrt N}),
\end{eqnarray}
 where $q$ and $p$ are now
dimensionful quantities and \mw \ has units of length over momentum
and has the physical interpretation of the ratio of the natural
distance and momenta scales of the problem (see
appendix~\ref{app:scales}).  In these coordinates the symplectic
two-form is in standard Darboux form $\omega = d p_{\mu} \wedge d
q_{\mu}$, or equivalently $\{ q, p\}_{\rm PB} = 1$ .

We have been extremely explicit in order to contrast our choice with
those of the literature.  For instance in the coordinates of
[\ref{ROOS}] $\{\tilde q, \tilde p\}_{\rm PB} = N$.  This point is
obfuscated because the correct equations of motion are maintained by
dividing the classical action (and hence $\omega$) by $N$.  Such a
procedure, however, leads to improper quantization.  Similarly in
terms of the projective coordinates [\ref{OriginalGilmore}] used to establish
algebraic-geometric correspondences in
such references as [\ref{AMADO},\ref{IA}] one has $\{ \tilde q, \tilde
p\}_{\rm PB} = N(1+ \tilde q^2 +\tilde p^2)^2$.  Thus only near the
domain $\tilde p^2 + \tilde q^2 \approx 0$ is it suitable to interpret these
coordinates as the natural coordinates of a cotangent bundle over
configuration space.

Classical Hamiltonians obtained from our coordinates usually have contributions
such as $p_r^4$ and $p_r^2 \, r^2$.  This leads to canonical
quantization ordering ambiguities as well as higher degree
differential equations.  However, if we focus on the lowest bound states we may use an approximate hamiltonian
which is valid in the classical regions whose paths most greatly
contribute to these states.  In the spirit of the stationary phase
approximation, the appropriate region would be near the fixed point of
the hamiltonian flow, i.e. where $dH = 0$.  

Fortunately, since the CSPS limit of algebraic hamiltonians typically have a convenient momenta dependence, i.e. $H = p_r^2 f(p_r^2, r^2) + V_{\rm eff}(r) $,
this condition implies we are looking in the region of reduced phase space near $p_r = 0$ and $r = r^*$ where $V_{\rm eff}'(r^*)=0$.  These conditions are eminently reasonable.

Near this region the dynamics can be given by approximating the hamiltonian:
\begin{eqnarray}
\label{eqn:harmonichamiltonian}
H(p_r,r) \approx 
 {H|}_{p_r=0, r=r^*} +\frac{1}{2} \left( V''_{\rm eff}(r^*) (r-r^*)^2 +
                   {H_{1^2}|}_{p=0, r=r^*}p_r^2      \right)  ,
\end{eqnarray}
where all other lower order terms vanish. Hence, at least locally
around the fixed point of the system, our ordering ambiguities are
resolved.  If $r^*$ is sufficiently far from the $r=0$ phase space
boundary one can sensically re-quantize.

Given our coordinate choice  each higher derivative in
the taylor expansion of $H$ in equation~\ref{eqn:harmonichamiltonian}
will be down by a factor of $\frac{1}{\sqrt N}$. Thus this
approximation may be considered as an expansion in large $N$.  Note,
however that this procedure is distinct from other `large $N$' techniques
such as the contraction limit [\ref{GILMOREBOOK}, \ref{IL}].  Indeed, we will see in section \ref{sec:U2requantization} that this procedure, in the specific instance for Hamiltonians with the dynamical symmetry $U(n) \supset U(n-1)$, essentially reproduces the contraction limit results.  

We have carried out this discussion in a reduced phase space with a radial
coordinate.  That is, due to $O(3)$ invariance the fixed point is not a true minima in the global phase space.  In instances where one has a true fixed point in the global phase space (e.g. when the minima is at $r=0$) a similar procedure applies.

The introduction of the scales in equation \ref{eqn:alphascale}
appears arbitray.  However, since we desire a generic correspondence which
preserves the inner product structure of the algebra in the
Schr\"odinger picture, once a scale is picked for one Hamiltonian it
must be fixed for others, i.e.  any scale change would influence the
overlap of eigenstates which should be equated with the inner product
of the representation. Thus, the scale for each hamiltonian must be
functionally related to the scale of another.  If the relationship is
not trivial $\mw$ may depend on the parameters within
the hamiltonian.  This dependence of the scale ratio, \mw, will be
suppressed until such subtleties must be considered, at which time we
will supplement our notation with subscripts.

\section{An Explicit Example: $U(3)$}  \setcounter{equation} {0}
\subsection{Coherent State Limit} 
\label{sec:U3limit}

The algebraic approach for 2D problems was presented by [\ref{IO}].  One considers symmetric representations of $u(3)$ realized with the two chains of interest:
\begin{eqnarray}
\begin{array}{ll}
U(3)\supset U(2) \supset O(2) & {\rm I} \\
U(3)\supset O(3) \supset O(2) & {\rm II}
\end{array}.
\end{eqnarray}
We use the same notation as [\ref{IO}] for the generators.  However,
we select a different $O(3)$ subgroup which is generated by
$\hat{R}_+$, $\hat{R}_-$, and $\hat{l}$.  Our choice introduces more
`coordinate like' terms of the classical Hamiltonian.  The choice
makes no difference for spectra and only an overall phase for FC
overlaps [\ref{SBI}]. Please note that the $O(3)$ group is a dynamical
symmetry subgroup and does not have the interpretation of a rotation
in configuration space.

The general Hamiltonian of a $U(3)$ model is
\begin{equation}
\label{eqn:U3hamiltonian}
H = \epsilon \, \hat n + \delta  \,\hat n (\hat n +1)
    + \beta \, \hat l^2 - A \, \hat W^2 ,
\end{equation}
 where $\epsilon$, $\delta$, and $A$ are taken as positive or $0$.  Setting $A=0$ 
 ($\epsilon = \delta = 0$) gives a Hamiltonian with dynamical symmetry I (II).  The basis corresponding to chain I is labeled by the eigenvalues of  $\hat{n}$ and $\hat{l}$ respectively.  The basis corresponding to chain II is labeled by the  eigenvalues of $\hat{W}^2=\frac{1}{2}(\hat{R}_+\hat{R}_- +\hat{R}_-\hat{R}_+) + \hat{l}^2$ and again $\hat l$.  

 Taking the appropriate expectations one may calculate the classical limits of the various operators.  The results are enumerated in Table~1. 

We reduce the phase space in the natural way:
\begin{equation}
\begin{array}{cc}
q_x = r \cos \theta & q_y = r \sin \theta \\*
p_x = p_r \cos \theta - p_{\theta} \, r \sin \theta &
p_y = p_r \sin \theta + p_{\theta} \, r \cos \theta ,
\end{array}
\end{equation}
implying
\begin{equation}
q \cdot q  =  r^2, \ \ \  p \cdot p = p_r^2 + \frac{l_{\rm cl}^2}{r^2},
\end{equation}
where $l^2_{\rm cl} = p_{\theta}^2 r^4$ is a constant of the motion.

\subsection{$U(2)$ Chain}
\label{sec:U2requantization}

We work out the re-quantization procedure in the case where $A=\beta=\delta=0$.   In this instance we have the Hamiltonian on the reduced phase space:
\begin{eqnarray}
 H_{\rm cl} = \epsilon  \, 
 \frac{\mw}{2 \hbar} \left( \frac{1}{(\overline {m \omega})^2}  (p_r^2 + \frac{l_{\rm cl}^2}{r^2})   + r^2 \right) .
\end{eqnarray}
The canonical quantization scheme yields the following prescription:
\begin{eqnarray}
\label{eqn:quantization}
p_r^2 \mapsto - {\hbar^2 \over r} {d \over dr}
                \left( r  {d \over dr} (\cdot) \right), \;\;\;\;\;
l \mapsto \hbar l \, ,
\end{eqnarray}
where $l$ now labels irreps. of $SO(2)$.  This yields the quantum Hamiltonian:
\begin{eqnarray}
H=  \epsilon  \,
  \frac{ \hbar}{2 \mw}
 \left( 
- {1 \over r} {d \over dr} \left( r  {d \over dr} (\cdot) \right)
+ {l^2 \over r^2}
+  \frac{(\overline {m \omega})^2}{\hbar^2} r^2
 \right),
\end{eqnarray}
which we immediately recognize as a multiple of a $2D$ circular oscillator with energy levels spaced by $\Delta E= \epsilon$ [\ref{F}]. This result  perfectly agrees with direct algebraic evaluation.

We proceed similarly for $\delta$ different from $0$.
The classical Hamiltonian is then of the form:
\begin{equation}
H = N(\epsilon + 2 \delta) n_{\rm cl} + \delta N (N-1) ( n_{\rm cl})^2.
\end{equation}
In Cartesian coordinates the conditions for a fixed point are:
\begin{equation}
\begin{array}{l}
dH = 0 = N [ \epsilon + 2 \delta + 2 \delta (N-1) n_{\rm cl} ] \\
\; \; \; \; \; \; \; \; \; \; \; \; \; \; \; \; \; \; \; \; \; \; \; \; 
\; \; \; \; \; \; \; \; \; \; \; \; 
       \times \left(
                \frac{\partial  n_{\rm cl}}{\partial x} dx +
                \frac{\partial  n_{\rm cl}}{\partial y} dy +
                \frac{\partial  n_{\rm cl}}{\partial p_x} dp_x +
                \frac{\partial  n_{\rm cl}}{\partial p_y} dp_y 
                \right),
\end{array}
\end{equation}
which is true at $x=y=p_x = p_y=0$ (hence our use of Cartesian as opposed to radial coordinates). Near this point to second order our Hamiltonian behaves like:
\begin{equation}
H \approx  N(\epsilon + 2 \delta) n_{\rm cl}.
\end{equation}
Re-quantization proceeds exactly as in the previous case leading to an
energy spacing of $\Delta E \approx \epsilon + 2 \delta$.
The corresponding algebraic Hamiltonian ($H = \epsilon \hat n + \delta \hat n (\hat n +1)$) has spacing for the lowest levels of $\Delta E = \epsilon + 2 \delta$ .

Additionally the same low energy eigenfunctions diagonalize both $U(2)$ Hamiltonians in both the algebraic and approximate re-quantizing prescriptions (i.e. in both prescriptions the inner-product of the lowest eigenfunctions of the Hamiltonain with $\delta = 0$ and $\delta \neq 0$ is $1$).

\subsection{Radially Displaced Oscillators---Near the $O(3)$ Chain}

In the situation where our Hamiltonian is near the $O(3)$ dynamical symmetry the fixed point of the Hamiltonian flow is not at $r=0$.  

The potential is given by:
\begin{equation}
\begin{array}{c}
V_{\rm eff}(r)
=  \frac{\mw}{\hbar}
 \left[ (X- Z ) \, \frac{1}{2} r^2 + (Y+Z) \frac{\mw}{4 N \hbar} r^4 \right. \; \; \; \; \; \; \; \;
 \\ \left.   \; \; \; \; \; \; \; \; \; \; \;  \; \; \;
            \; \; \; \; \; \; \; \; \; \; \; \;   \; \; \; \; \; \;  \; \; 
           \; \; \; \; \; \; \; \; \; \; \; \; \; \; \; \; \; \; \; +
X \, \frac{1}{2} \frac{1}{(\overline {m \omega})^2}\frac{l^2}{r^2}+
 Y  \frac{1}{4 N \hbar ({\overline{m \omega}})^3}\frac{l^4}{r^4} \right] ,
\end{array}
\end{equation}
where we have made the definitions $X= \epsilon + 2 \delta + \beta $, $Y=\delta  \, (N-1)$, and $Z=4 A \, (N-1)$.
The physically interesting regime [\ref{ME}] is `near' the $O(3)$ chain, i.e. $X\over Z$ and $Y \over Z$ are both small.  In this limit it is easy to find a perturbative solution for the condition $V'_{\rm eff}(r^*) =0$ :
\begin{equation}
\label{eqn:displace}
(r^*)^2  
\approx \frac{N \hbar}{\mw } 
  \left( 1 +{\frac{X}{Z}
              ( \frac{ l^2}{(N \hbar)^2}  -1) - \frac{Y}{Z} } 
   \right),
\end{equation}
where now it is clear that the presumption that $r^*$ is large depends on $N$ being large. Note that having $X$ or $Y$ different from $0$ decreases $r^*$.  That is, the $O(3)$ chain represents the maximum possible displacement.  $r^*$ may increase with increasing $l$ however, corresponding to a larger angular momentum barrier. 

Since $l$ can be as large as $N$ we regard $l^2/N^2$ as an independent quantity---even though in practice it is of the same order as terms which go $1/N^2$.  In anticipation of this we dropped the term going with $\frac{l^2}{N^4}$ for simplicity. 

When one moves slightly off the $O(3)$ chain different $l$ subspaces
which were degenerate split resulting in the lowest energy difference,
$\Delta E$, having a slight $l$ dependence.  As we've seen this
dependence has entered in lowest order in the form of terms
$\frac{l^2}{N^2}$. Since the ratio $\frac{l}{N}$ can be regarded as a
separate independent small quantity ignoring terms like
$O(\frac{1}{\sqrt N})$ does not mean we loose all $l$ dependence in
our calculation.  Our results for the energy level spacing will only
be correct to leading order in $N$.  The $l$ space splitting behavior
will be given correctly---but the actual values of the splitting will
be insignificant compared to other (non-$l$ dependent) contributions
we have ignored for smaller $l$.

Continuing we see:
\begin{eqnarray}
V''_{\rm eff}(r^*)
\approx  2 Z \frac{\overline{m \omega}}{\hbar}
 \left[ 1+\frac{X}{Z} 
           \left(3 \frac{ l^2}{(N \hbar)^2}-1 \right)   
       \right] , \\
H_{p_r^2}|_{p_r=0, r=r^*}   
   \approx
   \frac{Z }{2 \hbar (\overline{m\omega } )} \left[ 
     1 +{\frac{X}{Z}
              ( \frac{ l^2}{(N \hbar)^2}  +1) +
             \frac{Y}{Z}
                          (2 \frac{l^2}{(N \hbar)^2} +1) } 
   \right].
\end{eqnarray}

Near the fixed point the Hamiltonian's behavior is given by equation~\ref{eqn:harmonichamiltonian}.  However, the radial quantum mechanics depends crucially on the point $r=0$.  Since equation \ref{eqn:harmonichamiltonian} does not reproduce the original Hamiltonian's behavior near this point the approximation becomes suspect.  It would be more reasonable to quantize:
\begin{equation}
                   \frac{1}{2} {H_{{p_r}^2}|}_{p=0, r=r^*} p_r^2       
                                  +  \tilde V(r),
\end{equation}
where $\tilde V(r)$ has the same behavior near the fixed point as the approximate Hamiltonian and the behavior near $r=0$ of the original Hamiltonian.  That is:
\begin{eqnarray}
\lim_{r \rightarrow 0} r^4 \tilde V(r) & = &
   \frac{Y}{4} \frac{l^4}{N^2 \hbar ^2(\overline {m \omega})^2} \\
\tilde V(r) & = & {H|}_{p_r=0, r=r^*} +\frac{1}{2}  V''_{\rm eff}(r^*) (r-r^*)^2 +O((r-r^*)^3).
\end{eqnarray}
Additionally, of course, $\tilde V$ should not introduce any other minima at other locations. 

Following the prescription for the harmonic oscillator we use the radial quantization scheme (equation~\ref{eqn:quantization}),
 substituting for the wavefunction, $\Psi = {\psi \over \sqrt r}$, we obtain:
\begin{equation}
\label{eqn:psiODE}
- H_{{p_r}^2} {\hbar^2 \over 2 } \psi '' + 
\left[ \tilde V(r) - {H_{{p_r}^2} \hbar^2 \over 8 r^2} \right] \psi = E \psi ,
\end{equation}
where $\psi(0)=0$ and $\psi(r\rightarrow \infty)=0$.  The form of this equation is that of a 1-D Schr\"odinger equation except the left boundary condition is applied at $r=0$.    At this point the $r\approx 0$ dependence of $\tilde V$ becomes crucial.  Since $\tilde V$ blows up more quickly than the counter term blows down the sum of the two still has a minima, although it should be shifted inward from $r^*$.   Calling the new minima $r^{**} = r^*(1 - \epsilon)$ and using our knowledge of the behavior of $\tilde V$ near $r^*$ we find that the condition $\left. {d\over dr}\right|_{r= r^{**}}\left[ \tilde V(r) - {H_{{p_r}^2} \hbar^2 \over 8 r^2} \right] = 0$ yields:
\begin{equation}
\label{eqn:epsilon}
\epsilon \approx {1 \over16 N^2 -3}
              \left[ 1- {16 N^2  \over 16 N^2 -3}
                                    \left(4 \frac{X}{Z} (\frac{l^2}{N^2} -1) -
                                                         \frac{Y}{Z} (2 \frac{l^2}{N^2} +3)
                                                \right)
                          \right]
\end{equation}
This can be expanded in $N$ with leading contribution $O(\frac{1}{N^2})$.  Recalling that $N$ essentially counts the number of bound states (hence effectively measures the depth of the potential) it is reasonable that $1\over N^2$ should parameterize $\epsilon$.  Of course, since we are only working in leading order in $N$ we have $\epsilon \approx 0$ and consequently $r^{**} \approx r^*$.  Our primary intent of displaying equation~\ref{eqn:epsilon} is that the condition of $N$ being large is now clarified.  Since the prefactor must be small we find $2N \gg 1$.

Returning to equation \ref{eqn:psiODE} we notice that in our approximation  ($r^*$ or $r^{**}$ is large) a solution with the left boundary condition at $r = -\infty$ will, to a good approximation, satisfy the boundary condition at $r=0$.  In this instance the lower eigenvalues are given by $E \approx \hbar \omega$ where
\begin{equation}
\omega^2 = H_{{p_r}^2} \left[ \tilde V''(r^{**}) - 
                                   {3 H_{{p_r}^2} \hbar^2 \over 4 (r^{**})^4} 
                                          \right].
\end{equation}

Given the properties of $\tilde V$ we have $\tilde V''(r^{**})\approx \tilde V''(r^{*}) = V''_{\rm eff}(r^*)$.  The second term is $O(\frac{1}{N^2})$ and may be ignored.

Thus we have
\begin{equation}
\Delta E \approx {Z}   \left[ 
     1 +{2 \frac{X}{Z}  \frac{l^2}{N^2}
             + \frac{1}{2} \frac{Y}{Z} \left( 2 \frac{l^2}{N^2} +1 \right) 
                             }  \right].
\end{equation}
We see that on the $O(3)$ chain ($X=Y=0$) we have a spacing of 
$\Delta E \approx Z \approx 4AN$,
as compared to the exact algebraic expression:
\begin{equation}
\Delta E = A \left[ N (N+1) - (N-2)(N-1)  \right]= A(4N-2),
\end{equation}
which agrees to leading order in $N$ as advertised.

Next we compute the induced harmonic dilatation constant of the geometry (see Appendix \ref{app:scales}):
\begin{eqnarray}
\label{eqn:scale}
\alpha^2 \approx  { \omega \over \hbar H_{1^2} } 
\approx
2 \frac{ (\overline{m\omega } )}{ \hbar }
 \left[ 
     1 
          +{\frac{X}{Z}  \left(\frac{l^2}{N^2}-1 \right)
          - \frac{Y}{Z} \left(\frac{l^2}{N^2}  + \frac{1}{2} \right)
           }  \right] .
\end{eqnarray}
 Note, that the `concavity' corrections depend explicitly on the algebraic parameters, indicative of the fact that our calculations have established relationships between algebraic Hamiltonians and the geometry of configuration space. 

Equations \ref{eqn:scale} and \ref{eqn:displace} for the harmonic dilatation and radial displacement can be easily related to experimental data and provide useful assistance when analyzing transition intensities as to be detailed in the upcoming publication [\ref{ME}].

\subsection{Near $O(3)$ to $O(3)$ Inner Product}
\label{sec:nearO3FCfactors}

We wish to calculate the inner-product of eigenstates of  a hamiltonian corresponding to slightly off $O(3)$ symmetry to one on $O(3)$.  In either case the approximate wavefunctions are:
\begin{equation}
\psi_0^{\rm disp} \approx {\sqrt \alpha \over \pi ^{\frac{1}{4}}}
{e^{- \frac{1}{2} [ \alpha (r - r^*)]^2} \over \sqrt{r}}, 
\end{equation}
where we have normalized it on the interval $[{- \infty{\rm ,}\infty }]$ (as opposed to $[{0{\rm ,}\infty }]$) as this only introduces errors of order $\frac{1}{N}$.  The parameters $\alpha$ and $r^*$ are determined by \ref{eqn:scale} and \ref{eqn:displace}. We use $\alpha '$ and ${r^*}' $ as the parameters of the wavefunction exactly on the $O(3)$ chain (determined by the same equations except $X=Y=0$).   Note that there is an $l$ dependence hidden in both $r^*$ and $\alpha$.

Since both wavefunctions are essentially $0$ for $r < 0$ we may again calculate the overlap on the interval $[{- \infty{\rm ,}\infty }]$.  Thus we wish to evaluate the integral
\begin{equation}
I_{0,0} \approx \int_{- \infty}^{\infty} \phi(\alpha; r-r^*)
                                         \phi(\alpha'; r-{r^*}') dr,
\end{equation}
where $\phi$ is the ground state wavefunction of a 1D SHO.  The integral is easily evaluated to [\ref{II}]:
\begin{equation}
I_{0,0} \approx 
                   \left[{2 \alpha \alpha' \over \alpha^2 + \alpha'^2 }
                   \right] ^ {\frac{1}{2}} \times
                    \exp \left[- {(\alpha \alpha' ({r^*}'-r^*))^2 \over
                              2 ( \alpha^2 + \alpha'^2)} \right].
\end{equation}
As stated previously the ratio of scales $\overline{ m \omega}$ need not be the same in both scenarios and may indeed depend on the value of $X$,$Y$, and $Z$.  However, this dependance must be smooth.  Since the difference between these two scenarios is perturbative one concludes that ${\overline{ m \omega}\over\overline{ m \omega}'} = 1+ \delta $  where $\delta \sim O(X/Z,Y/Z)$.  Given this,
we have: 
\begin{equation}
\label{eqn:FC1}
I_{0,0} \approx 1 + O\left( (\frac{X}{Z})^2,(\frac{Y}{Z})^2 \right)
\end{equation}
This is the exact overlap one would expect from perturbation theory of the algebraic model as spelled out in Appendix B.  

\subsection{$O(3)$ to $U(2)$ Inner-Products}
Given equation~\ref{eqn:FC1}, to our order the overlaps from a hamiltonian near $O(3)$ to the $U(2)$ chain should be exactly the same as the exact $O(3)$-$U(2)$ factors. 

For the $U(2)$ chain we have the ground state wavefunction [\ref{F}]:
\begin{equation}
\label{eqn:U2groundstate}
\psi_0^{U(2)} = \sqrt{\frac{2}{|l|!}} (\alpha_{U(2)})^{|l| +1} r^{|l|} e^{- \frac{1}{2} [ \alpha_{U(2)} r]^2},
\end{equation}
where $(\alpha_{U(2)})^2 = \frac{\overline{m \omega}_{U(2)} }{\hbar}$.

Thus our overlap becomes:
\begin{eqnarray}
\label{eqn:overlapintegral}
\lefteqn{
I_{0,0}({\alpha_{U(2)} \over \alpha},\alpha r^*) = 
\sqrt{\frac{2 \alpha}{|l|!}} 
\frac{(\alpha_{U(2)})^{|l| +1}}{\pi ^{\frac{1}{4}}} 
\exp \left[ -\frac{1}{2}{ \alpha^2 \alpha_{U(2)}^2 \over {\alpha_+}^2 } 
            {r^*}^2 \right]}  
                            \\ \nonumber & & \; \; \; \; \; \; \;
\; \; \; \; \; \; \; \; \; \; \; \; \; \; \; \; \; \; \; \; \; \times 
\int_0^\infty  dr \;
  r^{|l|+\frac{1}{2}} 
  \exp \left[{- \frac{1}{2} ( {\alpha_+} \, r 
                   - \frac{\alpha^2 r^*}{\alpha_+})^2}
        \right]
\end{eqnarray}
where $\alpha_+^2 = {\alpha_{U(2)}^2+\alpha^2}$, and $\alpha$ is again defined by \ref{eqn:scale} with $X=Y=0$ .  The integral may be expressed in terms of confluent hypergeometric functions:
\begin{eqnarray}
\nonumber \lefteqn{
\int_0^\infty  dr \;
  r^{|l|+\frac{1}{2}} 
  \exp \left[{- \frac{1}{2} ( {\alpha_+} \, r 
                   - \frac{\alpha^2 r^*}{\alpha_+})^2} \right] 
\; \;   =       
2^{\frac{2l-1}{4}} \frac{1}{{\alpha_+}^{\frac{3 + 2l}{2}}}
\exp \left[{-\frac{1}{2} \left(\frac{\alpha^2 r^*}{\alpha_+}\right)^2} \right]
} \\ & & \times
\left\{
\Gamma\left(\frac{3+2l}{4}\right) 
\ _1 F_1\left( \frac{3+2l}{4}, \frac{1}{2}; 
               {\frac{1}{2} \left(\frac{\alpha^2 r^*}{\alpha_+}\right)^2} \right) 
\right. \\ \nonumber  & &  
 \;\; \; \; \; \;\; \; \; \; \;\; \; \; \; \; \left.
+ \sqrt{2}\left(\frac{\alpha^2 r^*}{\alpha_+}\right)
 \Gamma \left( \frac{5+2l}{4} \right)
 \ _1 F_1\left( \frac{5+2l}{4}, \frac{1}{2}; 
                {\frac{1}{2} \left(\frac{\alpha^2 r^*}{\alpha_+}\right)^2} \right)
\right\}.
\end{eqnarray} 
We note that $\frac{\alpha^2 r^*}{\alpha_+} \sim N$ and hence use an
asymptotic expansion for the hypergeometric functions [\ref{ASi}].
Keeping only the most dominant contribution of the integral we find:
\begin{equation}
I_{0,0} 
\approx 
\frac{\pi ^{\frac{1}{4}} }{\sqrt{|l|!}} 
\left( \frac{\alpha_{U(2)}}{\alpha} \right)^{|l| +1} 
 \left( \frac{\alpha}{\alpha_+} \right)^{4+2|l| } 
 \exp \left[ -\frac{1}{2} { \alpha^2 \alpha_{U(2)}^2 \over {\alpha_+}^2 } 
            {r^*}^2 \right]
(\alpha  r^*)^{\frac{5+2|l| }{2} }
.
\end{equation}
Substituting the expressions for the harmonic scales, and letting $\zeta = 
\frac{\overline{m \omega}_{U(2)}}{\overline{m \omega}_{O(3)}}$ we find
\begin{equation}
\label{eqn:O3U2overlap}
I_{0,0}
\approx 
\frac{\pi ^{\frac{1}{4}} }{\sqrt{|l|!}} 
\left( \frac{\zeta}{2} \right)^{|l| +1} 
 \left( \frac{2}{\zeta +2} \right)^{2+|l| } 
 \exp \left[ -N {1 \over 1 + \frac{2}{\zeta} } 
             \right]
(2 N)^{\frac{5+2|l| }{4} }
.
\end{equation}

\subsection{Scale Changes}

Comparing the analytic expression of the $U(2)$/$O(3)$ overlaps (equation~\ref{eqn:O3U2overlap} calculated by approximately requantizing) and the expression from appendix equation~\ref{eqn:O3U2overlapalgebraic} (calculated by direct algebraic means) fixes the scale dependance.  Comparing the exponential dependance of the two expressions,
\begin{equation}
e^{- \frac{\ln \, 2}{2} N} \; {\rm vs.} \; 
e^{\left[ -N {1 \over 1 + \frac{2}{\zeta} } 
             \right]},
\end{equation}
leads to the conclusion that:
\begin{equation}
\zeta = 
\frac{\overline{m \omega}_{U(2)}}{\overline{m \omega}_{O(3)}}
\approx
{2 \ln 2 \over 2 - \ln 2}.
\end{equation}

The rest of the dependance can be fixed by allowing $\zeta$ to have $\log N \over N$  and $1 \over N$ corrections.  To see this we let:
\begin{equation}
\zeta \approx {2 \log 2 \over 2 - \log 2} 
                \left( 1 - \frac{4}{\log 2 (2- \log 2)} \frac{\log \gamma}{N}
                                \right),
\end{equation}
or equivalently
\begin{equation}
(1 + \frac{2}{\zeta})^{-1} \approx \frac{ \log 2}{2} - \frac{\log \gamma}{N}.
\end{equation}
Substituting into the expression~\ref{eqn:O3U2overlap} we find the first order term of the large $N$ asymptotic:
\begin{equation}
I_{0,0} \approx \frac{\pi^{\frac{1}{4}}}{\sqrt{|l|!}} 
                \left[ {\log 2 \over 2} \right] ^{|l|+1}
                                \left[ {2-\log 2 \over 2} \right]
                                (2N)^{\frac{5+2|l|}{4}}
                                e^{-N {\log 2 \over 2}} \gamma.
\end{equation}

Comparing with the algebraic expression we conclude that
\begin{equation}
\gamma = \frac{2^{\frac{3}{4}}}{2 - \log 2}
          (\log 2)^{- (|l| +1)} N^{- \frac{3}{2}}
\end{equation}

That is, in order for the Schr\"odinger and algebraic prescriptions to be commensurate for the leading asymptotic in $N$, we must have the first orders of the asymptotics of $\zeta$:
\begin{eqnarray}
\label{eqn:scaledependance}
\zeta  \approx {2 \log 2 \over 2 - \log 2} 
                \left[ 1 + \frac{4}{\log 2 (2- \log 2)} 
                                     \left\{ \frac{3}{2} \frac{\log N}{N} +
                                            (|l| \log{ \log 2} + c) \frac{1}{N}
                                     \right\}
                                \right] ,  \\
  c  = \log \log 2  - \log \frac{2^\frac{3}{4}}{2 - \log 2} \; \; \; \;
  \; \; \; \; \;  \; \; \; \; \;   \; \; \; \; \;  \; \; \; \; \;   \; \; \; \; \;  \;  \; \;  \; \; \; \; \; 
\end{eqnarray}

Calculating the inner product of the eigenstates of an algebraic
hamiltonian on the $U(2)$ chain with the eigenstates of another
hamiltonian off of the $U(2)$ chain is not simply analogous to the
overlap of radially displaced oscillators, but analagous to the matrix
elements of an operator which radially displaces {\em and} dilatates
(much like the operator matrix elements calculated in [\ref{II}]) changing
the natural scale of the problem.
The degree of dilatation depends on the proximity of the second
hamiltonian to either chain.  For chains near $U(2)$ the dilatation
paramater is essentially $1$.  As one moves nearer to the $O(3)$ chain
the dilatation parameter increases, approaching the value given by
equation \ref{eqn:scaledependance}.  This has considerable
consequences for hybrid algebraic-Schr\"odinger analysis of molecules
[\ref{ME}].

\section{Conclusions} \setcounter{equation} {0}
We have provided a procedure, via requantization, to convert a hamiltonian of an algebraic model into a Schr\"odinger hamiltonian which will give the same results for the lower states to leading order in $N$.  The procedure is optimized for algebras which are interpreted as single particle hamiltonians.  Thus, it seems most applicable to molecular models such as the vibron model $U(4)$ [\ref{U4}], the anharmonic oscillator $U(2)$ [\ref{MANY}], and the two dimensional $U(3)$ model considered here.  Although the prescription should generate Schr\"odinger hamiltonians in other models it remains to be seen whether or not such results would have appropriate many-body interpretations.

We have carried out the requantization process in detail for the limiting cases of the $U(3)$ model.  By demanding that the wavefunction overlap of the requantized system agree with the inner product of the algebraic model we have demonstrated that different chains of the $U(3)$ model correspond to not only different geometries but different scales.   

\section{Acknowledgements} \setcounter{equation} {0}
This work was performed in part under the U.S. Department of Energy, Contract No.~DE-FG02-91ER40608.  I extend my deepest gratitude to my advisor, Prof.~Franco Iachello, for introducing me to the problem and his suggestions and feedback on this work.  I thank Prof.~Dimitri Kusnezov for also making suggestions on this manuscript.  Finally, I would like to thank the Department of Energy's Institute for Nuclear Theory at the University of Washington for its hospitality during the completion of this work.


\

\

\renewcommand{\theequation} {{\Alph{section}}.{\arabic{equation}}}

\noindent
{\Huge {\bf Appendices}}
\appendix

\section{Scales in the Schr\"odinger Picture}  \setcounter{equation} {0}
\label{app:scales}

 Consider a traditional Schr\"odinger Hamiltonian (1-D for simplicity),
\begin{equation}
H= - \frac{\hbar^2}{2m}{d^2 \over dx^2} + V_0 f(\bar \alpha x),
\end{equation}
 with the distance scale $\bar \alpha$ and momenta scale given by  $\frac{\bar \alpha}{\overline {m \omega}} = (m V_0)^{-\frac{1}{2}}$.  

We on occasion refer to the induced harmonic dilatation constant---which we define as the distance scale (dilatation constant of the wavefunctions) one obtains by approximating the potential about a minimum to second order.  For a SHO $f(x) = \frac{1}{2} x^2$ and $V_0 = {m \omega^2 \over \bar \alpha^2}$.  In this instance $\overline {m \omega} = m \omega$ and we define the distance scale $\bar \alpha^2 = {m \omega \over \hbar}$. If $x^*$ denotes the minimum of a more complicated potential we see that the distance scale associated with its harmonic approximation is 
\begin{equation}
\label{eqn:InducedScaleFormula}
\alpha^2 = {\overline {m \omega} \over \hbar} \sqrt{ f''(\bar \alpha x^*)}.
\end{equation}
The induced scale depends on two quantities: (1) the ratio of the true distance and momenta scales; (2) the concavity of the potential about the minima.

\section{Algebraic Calculations of Overlaps near $O(3)$} \setcounter{equation} {0}
\label{app:algebraic}

We wish to find the overlaps of eigenstates of the hamiltonian
\begin{equation}
H = A \left( {(\epsilon + \delta)\over A} \hat n + \frac{\delta}{A} \hat n (\hat n+1) + \frac{\beta}{A} \hat l^2 - \hat W^2  \right)
\end{equation}
to the hamiltonian with $\epsilon= \delta= \beta =0$ in the limit where $A$ is large.  We may work within subspaces of constant $l$ so that there are no degeneracies. 

We denote the basis from the $O(3)$ chain by $|[N],\omega,l \rangle$ where $\omega(\omega +1)$ is the eigenvalue of $\hat W^2$.  The $U(2)$ chain basis is denoted by $|[N],n,l \rangle$ analogously.  We denote the transformation matrix between them by $\zeta$, i.e. $|[N],\omega \rangle = \sum_n \zeta^\omega_n |[N],n\rangle$, where we have suppressed the implicit $l$ dependance.

The wavefunction for the ground state may be perturbatively calculated to first order:
\begin{equation}
\label{eqn:algperturbstate}
|[N],E_{\rm low} \rangle  \approx
|[N],\omega = N \rangle
+ \sum_{\omega' \neq N} |[N],\omega' \rangle
             {\sum_n (\zeta^{\omega '}_n)^* \zeta^{\omega=N}_n 
                         \left[(\epsilon + 2\delta) n + \delta n^2 \right] \over
 A N (N+1)-A \omega ' (\omega'+1)        }.
\end{equation}

As always in perturbation theory, to first order in  $\frac{1}{A}$ the wavefunctions only have corrections orthogonal to the unpeturbed wavefunction.  Consequently the overlap from a slight deformation off the $O(3)$ chain to the $O(3)$ chain is identically $1$ to this order. 

We now concentrate on the overlap between the $O(3)$ and $U(2)$ chains.  In this limit we have:
\begin{eqnarray}
I_{0,0} =\langle [N],n=|l|,l|[N],\omega = N,l \rangle =
\zeta^{\omega = N}_{n=|l|}
 .
\end{eqnarray}
 
 These coefficients are calculated in [\ref{SBI}]:
\begin{eqnarray}
\label{eqn:preO3U2overlapalgebraic}
\zeta^{\omega = N}_{n=|l|} =
\left[{(N+|l|)!
\over
 2^{|l|} \, |l|! \, (2N-1)!!
}\right]^{\frac{1}{2}}.
\end{eqnarray}

Using $(2N-1)!! = 2^N \Gamma(N+ \frac{1}{2})/ \Gamma(\frac{1}{2})$ and the Stirling approximation $\Gamma(z+b) \sim \sqrt{2 \pi}e^{-z} z^{z+b-\frac{1}{2}}$ [\ref{ASi}] this formula may be approximated for large $N$ by:
\begin{eqnarray}
\label{eqn:O3U2overlapalgebraic}
I_{0,0}  \approx
\frac{\pi^{\frac{1}{4}}}{2^{\frac{|l|}{2}}\sqrt{|l|!}}
N^{\frac{|l|}{2} -\frac{1}{4}} e^{- \frac{\log \, 2}{2} N}.
\end{eqnarray}

If one wishes to algebraically calculate FC transitions to higher energy $U(2)$ bound states one may proceed in the exact same fashion to find
\begin{eqnarray}
\label{eqn:O3U2overlapalgebraichigher}
I_{n,0}  \approx
{\pi^{\frac{1}{4}}  
N^{\frac{n}{2} -\frac{1}{4}} e^{- \frac{\log \, 2}{2} N}
\over 2^{\frac{n}{2}} \sqrt{(\frac{n+l}{2})! \, (\frac{n-l}{2})!}}.
\end{eqnarray}
%

%
%
%
%

\newpage

\begin{table}[hp]
\centerline{ 
\begin{tabular}{|clcc|}   \hline \hline
         & & Group Coordinates                                  & Projective Coordinates \\  [5 pt]
\cline{3-4} 
$\langle \hat l \rangle /N$
& \multicolumn{1}{l|}{$ \equiv l_{\rm cl}\ (\tilde l_{\rm cl})$}
& $q_x p_y - q_y p_x$   
& $2 \Sigma ({\tilde q_x \tilde p_y - \tilde q_y \tilde p_x })$ \\ [5 pt]
$\langle \hat n \rangle /N$
& \multicolumn{1}{l|}{$ \equiv n_{\rm cl} \ (\tilde n_{\rm cl}) $}     
& $\frac{1}{2} ( q \cdot q + p \cdot p)$                
& $\Sigma ( \tilde q \cdot \tilde q + \tilde p \cdot \tilde p) $ \\ [5 pt]
$\langle \hat n^2 \rangle /N$
& \multicolumn{1}{l|}{\ }       
& $n_{\rm cl} + (N-1) (n_{\rm cl})^2$           
& $\tilde  n_{\rm cl} + (N-1) (\tilde  n_{\rm cl})^2$ \\ [5 pt]
$\langle \hat l^2 \rangle /N$
& \multicolumn{1}{l|}{\ }       
& $n_{\rm cl} + (N-1) (l_{\rm cl})^2$           
& $\tilde  n_{\rm cl} + (N-1) (\tilde l_{\rm cl})^2$ \\ [5 pt]
$\langle \hat W^2 \rangle /N(N-1) $
& \multicolumn{1}{l|}{\ }       
& $ (l_{\rm cl})^2 +2(1-n_{\rm cl})  \,  q \cdot q  +\frac{2}{N-1}$             
& $ (\tilde l_{\rm cl})^2 +4 \Sigma^2 \, {\tilde q \cdot \tilde q }  +
            \frac{2}{N-1}(\Sigma + \tilde n_{\rm cl})
  $ \\ [5 pt]
\hline \hline
\end{tabular} }
%
%
\caption {The coherent state limit of $U(3)$ operators in group and projective coordinates. In group coordinates we take the expectation in the state $\frac{1}{\sqrt{N!}}( \sqrt{1-\alpha \cdot \alpha^*} \sigma^{\dagger} + \alpha \cdot \tau^{\dagger})^N|0 \rangle $ and let $\alpha =  \frac{1}{ \sqrt{2}} (q + ip)$ to reproduce the standard form of the angular momentum. In projective coordinates the unnormalized coherent state is parameterized by $\frac{1}{\sqrt{N!}}(  \sigma^{\dagger} + \tilde \alpha \cdot \tau^{\dagger})^N|0 \rangle $.  One must then divide the expectation by the normalization.  We then let $\tilde \alpha =(\tilde q + i\tilde p)$ to jibe with the standard definitions in the literature.  Note that the `$q$'s' and `$p$'s' in each column are not the same.  We have further made the abbreviation for projective coordinates: $\Sigma = {1 \over 1 +( \tilde q \cdot \tilde q + \tilde p \cdot \tilde p)} $.}
\end{table}

\end{document}